\documentclass[doublecol]{epl2} 
\usepackage{amsmath}

\title{Aging and relaxation near Random Pinning Glass Transitions}

\author{Chiara Cammarota \and Giulio Biroli}

\institute{                    
  Institut Physique Th\'eorique (IPhT) CEA Saclay, and CNRS URA 2306, 91191 Gif Sur Yvette, France
}

\abstract{
Pinning particles at random in supercooled liquids is a promising route to make substantial 
progress on the glass transition problem.
Here we develop a mean-field theory by studying the equilibrium and non-equilibrium dynamics of the spherical $p$-spin model in presence of a fraction $c$ of pinned spins.
Our study shows the existence of two dynamic critical lines: one corresponding to usual Mode Coupling transitions
and the other one to dynamic spinodal transitions. Quenches in the portion of the $c-T$ phase diagram delimited by those
two lines leads to aging. By extending our results to finite dimensional systems
we predict non-interrupted aging only for quenches on the ideal glass transition line and two very different types of equilibrium relaxations for quenches below and above it. 
}

\begin{document}

\maketitle

\section{Introduction}
The slowing down of the dynamics upon cooling is an universal property of supercooled liquids.
In the most spectacular cases relaxation time 
increases by near 15 orders of magnitude in a pretty narrow temperature window 
while the structural properties of the system remain almost unchanged.
Since the beginning, a recurrent question in the 
literature has been whether the fast growth of the relaxation time is a purely 
dynamic phenomenon or it is governed by the approaching of a thermodynamic singularity 
(the ideal glass transition) at a finite temperature $T_K$.
The problem in resolving this issue is related to the very nature of the glass transition: 
the relaxation time increases so steeply that it is not possible to approach the putative transition 
very closely. Below a certain temperature, still rather far 
from $T_K$,  the system becomes dynamically arrested---a glass---and falls out of equilibrium. \\
A way to partially bypass these equilibration problems and find a method to perform stringent tests
on the ideal glass transition and its thermodynamic character has been proposed in~\cite{BC}. In this recent work we suggested that 
pinning a fraction $c$ of particles of a supercooled liquid in an equilibrium 
configuration induces an ideal glass transition,
the Random Pinning Glass Transition (RPGT), at temperatures fairly higher than $T_K$. 
Using the Random First Order Transition (RFOT) theory we derived the phase diagram in the 
$c-T$ plane, in particular a line of RPGTs starting from $T_K$, $c=0$ and ending at certain values
$T_h,c_h$. 
Actually, the procedure of pinning particles is a powerful probe for studying glassy systems also for 
other reasons. It has been proposed in \cite{KP,BB} and then used in ~\cite{BBCGV,BK,BC,Sausset,Gilles} as a tool for detecting 
the presence of medium range amorphous order in supercooled liquids.
Moreover, it plays a role in studies of glassy dynamics in porous media. In fact, a simplified way to 
represent the porous matrix consists in freezing particles of an equilibrated liquid. 
The protocol leading to RPGTs correspond to the so called Equilibrated Mixture procedure studied in \cite{KIM,KMS}.  \\
In simulations, either the ones aimed at studying amorphous order or the ones 
focusing on porous media, it was found that the presence of pinned particles always produces a 
significative slowing down in the dynamics of the remaining free particles of the 
system~\cite{KIM,KMS,BK}. This phenomenon turns out to be 
fully consistent with what is expected from our RFOT analysis~\cite{BC}:
the slowing down of the dynamics is a precursor of the RPGT. 
A complete dynamic picture including out of equilibrium dynamics is, however, still missing. 
Many related interesting questions remain open.  For example, what happens if one quenches
a random pinned liquid below its glass transition temperature?  Is the subsequent aging
dynamics the same one of unpinned systems? \\
By analyzing the dynamics and the off-equilibrium dynamics of the randomly pinned p-spin spherical model, 
a system providing a mean-field theory of glassy systems, we shall answer these questions and 
show that indeed aging is quite different and that this is related to the very special nature of the ideal glass state for RPGT.
Moreover, we shall also obtain the Mode Coupling Transition (MCT) phase diagram
and critical properties, which differ from the ones obtained for porous media 
by using projection operators methods \cite{KRAKO}. A general discussion, in particular concerning the extension of our results 
beyond mean-field theory will be presented in the conclusion. The main finding of this work is that in finite dimensions
aging is always interrupted, except for quenches on the critical RPGT line. Surprisingly, aging is interrupted even for quenches {\it below} the RPGT line, but very differently than for quenches above it. In the former case the ideal glass transition state is suddenly nucleated 
from the slowly evolving aging state, whereas in the latter aging smoothly evolves into equilibrium dynamics. 
These surprising results can be tested in simulations or in experiments on colloids and provide new 
and interesting ways to ascertain the nature of the glass transition.

\section{Randomly pinned spherical $p$-spin model and dynamical protocols}
A mean-field study of the equilibrium and out of equilibrium dynamics of randomly pinned
systems can be obtained by focusing on the spherical $p$-spin model, which is known 
to provide a mean-field description of glassy dynamics.\\
The Hamiltonian of the model reads
\begin{equation}
H=-\sum_{i_1<\dots<i_p}^{N} J_{i_1\dots i_p} s_{i_1} \dots s_{i_p} \ .
\end{equation}
The $s_i$ are $N$ continuous variables satisfying the spherical constraint
$N=\sum_i^Ns_i^2$. The couplings $J_{i_1\dots i_p}$ are independent random variables 
extracted from a Gaussian distribution with zero mean and variance 
$p!/2N^{p-1}$. \\
At time $t=0$,  we pick a configuration $\overline{\mathcal{C}}$ at random from the equilibrium measure
and block a fraction $c$ of its spins: $\forall t\geq0$,
$s_i(t)=\overline{s_i}$ for $i\in[1,cN]$.
We focus on two different types of dynamics for $t\ge0$.
{\it Equilibrium dynamics:} 
the remaining free spins evolve directly from the values they take in the pinned configuration 
$\overline{\mathcal{C}}$: 
for $i\in[cN,N]$, $s_i(t=0)=\overline{s_i}$.
In this case the system starts from (and remains at) equilibrium.
{\it Quench dynamics and relaxation toward equilibrium:} 
the initial conditions for the remaining free spins are completely random, 
{\it i.e.} they assume the values of an infinite temperature configuration 
$\mathcal{C}^{\infty}$: for $i\in[cN,N]$, $s_i(t=0)=s_i^{\infty}$.
In this case we analyze the relaxation toward equilibrium and, when this does not take place, the corresponding
out of equilibrium aging dynamics.\\
In both cases for $t>0$ the free spins evolve according to the Langevin equation: for $i\in[cN,N]$
\begin{equation}
\dot s_i(t)=-\frac{\partial H}{\partial s_i}-\mu(t) s_i(t) + \eta_i(t) \ .
\end{equation}
The Lagrange multiplier $\mu(t)$ is introduced to enforce the spherical constraint and 
$\eta_i(t)$ is a Gaussian white noise with zero mean and variance $2T$.
We shall be interested in observables averaged over the possible choices of the couplings, the initial configuration 
$\overline{\mathcal{C}}$ and the
Gaussian noise.
\section{Equilibrium dynamics}
By using the method developed by Barrat, Burioni and M\'ezard~\cite{BBM} to study the equilibrium dynamics
we have obtained in the large $N$ limit the equation verified  by the correlation function 
$C(t,t^{\prime})=\frac{1}{N(1-c)}\sum_{i=cN}^{N}\langle s_i(t) s_i(t^{\prime})\rangle$. 
Since the system is time translation invariant (it is at equilibrium), 
the equation can be directly written on $C(\tau)=C(t,t^{\prime})$ where $\tau=t-t^\prime>0$:
\begin{align}
\partial_{\tau}C(\tau)=-TC(\tau)-\beta\int_0^{\tau}du \partial_uC(u)\mathcal{V}^{\prime}(C(\tau-u))
\label{EQCtau}
\end{align}
The kernel in the integral is the derivative of 
$\mathcal{V}(C)=\frac{1}{2(1-c)}[c+(1-c)C]^p$.
\begin{figure}
\onefigure[width=6.cm, angle=-90]{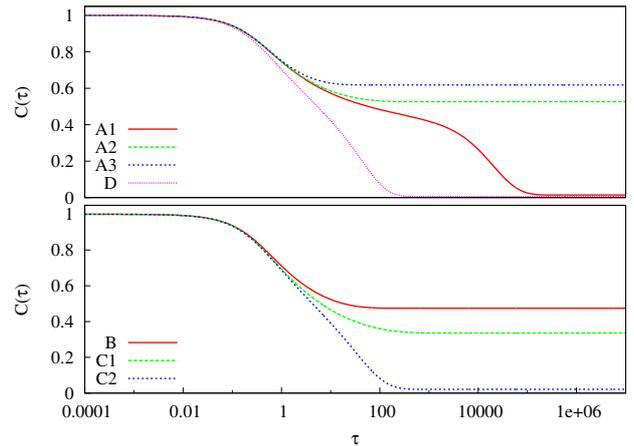}
\caption{Solutions of \eqref{EQCtau} for different values of the parameters $c$ and $T$. The letter code refers 
the positions in the $c-T$ phase diagram of Fig.\ref{PhDi}. The top panel shows the development of the plateau in the 
correlation function in proximity of the dynamical transition at $T_d(c)$. The bottom panel shows that the large time value of the correlation function decreases without jumps on a path around $(c_h,T_h)$.}
\label{EQdyn}
\end{figure}  \\
The solution of this equation has a shape similar to the one of the usual correlation function, see Fig.\ref{EQdyn}.
For small fractions of pinned spins and/or high temperatures (curves C2 and D in Fig.\ref{EQdyn} 
and corresponding spots in the phase diagram in Fig.\ref{PhDi}) the correlation function simply decays 
to a small value $q_0(c,T)\propto c$. 
Increasing $c$ and decreasing the temperature, the function develops an intermediate plateau and shows the usual 
two steps relaxation of the correlation function in proximity of the dynamic transition (curve A1).
The plateau appears at a finite value $q_1(c,T)$.
Finally, along a line in the $c-T$ plane that we denote either as $c_d(T)$ or $T_d(c)$, the correlation 
function ceases to depart from $q_1(c,T_d(c))$ on any finite time, {\it i.e.} the system undergoes 
a dynamic MCT transition. This is the continuation in the $(c,T)$ plane of the usual one 
that takes place at $T_d=\sqrt{\frac{p(p-2)^{p-2}}{2(p-1)^{p-1}}}$ and $c=0$.
The line of dynamical transitions was already shown in ~\cite{BC} as a result of the
static analysis. We plot it again in the more complete phase diagram\footnote{This phase diagram is similar and in 
agreement with the one found by Ricci-Tersenghi and Semerjian \cite{SR} in the context of combinatorial optimization problems.
Note however that the transition taking place at $T_{d_f}(c)$ is different from the one they discussed since we are focusing on non-equilibrium dynamics. } of Fig.\ref{PhDi}, where  
we have also drawn the thermodynamic transition line $T_K(c)$ discussed in \cite{BC}. 
The large time limit of~\eqref{EQCtau} gives an equation on 
$q=\lim_{\tau\rightarrow\infty}C(\tau)$: $q=\beta^2(1-q)\mathcal{V}^{\prime}(q)$
which for certain values of $c$ and $T$($\le T_d(c)$) has two solutions: $q_1(c,T)$ and $q_0(c,T)$. 
For these pairs $(c,T)$, $\lim_{\tau\rightarrow\infty}C(\tau)$ is given by the highest value $q_1$ ($>q_0$).
Note that we have checked that replica computations lead to the same equation for the self-overlap $q$.\\
The following picture results from the analysis of the dynamical equation.
Before the dynamic transition (points D, A1 in Fig.2) the system is in the paramagnetic-liquid phase. The small value of the correlation 
function at large time simply comes from the trivial influence of pinned spins towards their neighbors.
At the dynamical transition the ergodicity of the system is broken and the configuration of the free spins 
remains trapped in a small region of the phase space in proximity of the pinned configuration.
The large time value of the correlation function in this case corresponds to the self-overlap of the 
metastable state to which the pinned configuration belongs.
\begin{figure}
\onefigure[width=6.cm, angle=-90]{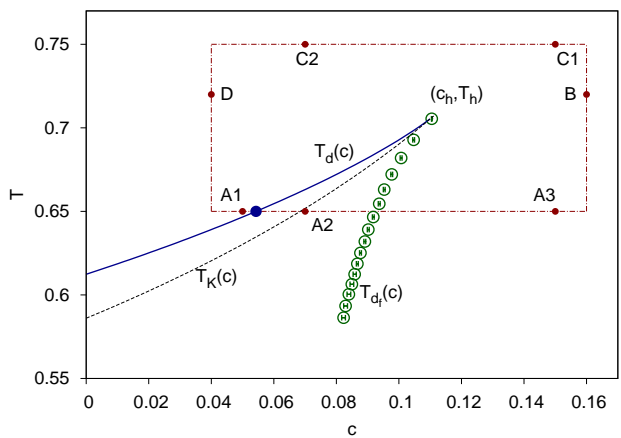}
\caption{Phase diagram of the $p$-spin model with pinned spins. The two dynamic transition lines meet with 
the thermodynamic transition line at $(c_h,T_h)$. Above this temperature no thermodynamic and dynamic 
transition is present in the system. The rectangle represents the closed path we followed to obtain curves 
of Fig.\ref{EQdyn}.}
\label{PhDi}
\end{figure}
Fig.2 shows that by increasing the temperature, a higher concentration $c_d(T)$ of frozen 
spins is required to induce the dynamical transition. Correspondingly, 
$q_0$ right before the transition increases. Instead, the self overlap of the metastable state in which the dynamics 
remains trapped lowers because the metastable states become broader. 
At the point $T_h=\sqrt{\frac{(p-2)^{p-2}}{2p^{p-3}}}$, $c_h=(p-2)^2/p^2$ 
at which $q_1$ reaches $q_0$, the dynamical transition line ends. This corresponds to a 
spinodal point for the metastable states, as it can be shown by the static replica computation.
Above $T_h$ the pinning procedure is not able to induce any ergodicity breaking: 
the high value of the large time correlation function is only due to trivial correlation between the free 
and the frozen spins. \\
The phase diagram in Fig.2 shows, surprisingly, that it is possible to follow a path in the $c-T$ plane that connects
the liquid and the ideal glass phase without crossing any phase transition, similarly to what happens for the 
liquid-gas transition. 
In order to illustrate this fact 
we discuss the behavior of the correlation function along the rectangular path drawn in Fig.\ref{PhDi}. 
$C(\tau)$ is singular only once, corresponding to the big spot in the figure. 
Afterwards, following the bottom side (A) of the rectangle the 
large time value $q$ of the correlation function increases with $c$ (see top panel of Fig. 1).
On the right and the top sides (B and C), $q$ decreases finally reaching, without showing any singularity, 
the low overlap solution $q_0$ of~\eqref{EQCtau}.
At the same time, approaching the transition from the left (last part of left side and A1), the plateau forms
anticipating the jump from $q_0$ to $q_1$ at the dynamical transition.\\
As a last issue, we address how the singular MCT behavior changes moving along the $T_d(c)$ line
in the phase diagram. The critical behavior is always the usual MCT one, called $A_2$ by G\"otze~\cite{GOTZE}.
The only difference is in the value of the
exponents $a$ and $b$ governing the approach and the departure of the correlation function from the plateau.
They are smaller for transitions occurring at higher concentrations of frozen spins.
Accordingly, transitions with higher critical temperatures are characterized by more stretched decays 
towards the $q_0$ large time limit.
The $a$ and $b$ exponent are fixed by the relation
\begin{equation}
\lambda=\frac{\Gamma^2[1-a]}{\Gamma[1-2a]}=\frac{\Gamma^2[1+b]}{\Gamma[1+2b]}=\frac{1}{2}\frac{\mathcal{V}^{\prime\prime\prime}(q_1(c,T_d(c)))}{[\mathcal{V}^{\prime\prime}(q_1(c,T_d(c)))]^{3/2}} \ .\nonumber
\end{equation}
For $c=0$ we recover the standard result $\lambda=0.5$, $a\simeq0.395$  and $b=1$. 
By increasing $c$, the values of the exponents decrease
and at the point $(c_h,T_h)$, at which $\lambda=1$,  they become identically zero. 
Correspondingly, at the end of the dynamical transition line, the correlation function shows a logarithmic decay 
characteristic of $A_3$ singularities, instead of the usual power law decay~\cite{GOTZE}.
Moreover, the relaxation time $\tau_{\alpha}$ diverges 
exponentially following a generalized Vogel-Fulcher-Tamman (VFT) law  $\tau_{\alpha}=\exp(D|T-T_h|^{-\theta})$
with exponent $\theta=1/6$ ~\cite{GOTZE}. 
It is remarkable that at the only point ($c_h,T_h$) where the dynamical transition does exist, since 
it coincides with the static one, mean-field theory correctly predicts a super-Arrhenius behavior. 
Finally, we point out that using the MCT theory of dynamical correlations \cite{BBMR} one finds that approaching
the endpoint $T_h,c_h$ the system becomes less dynamically heterogeneous since the growth of $\chi_4$ in time,
given by power laws with exponents $a$ and $b$, diminishes. Moreover, the peak of  $\chi_4$ at $c_h,T_h$
diverges less rapidly.

\section{Non-equilibrium dynamics}
The set of equations describing the non-equilibrium dynamics is obtained as in \cite{CK}; we have, for $t>t^{\prime}$,
\begin{subequations}
\label{NEeq}
\begin{align}
\partial_t C&(t,t^{\prime})\hspace{-0.05cm}+\hspace{-0.05cm}\mu(t)C(t,t^{\prime})\hspace{-0.05cm}= 
\hspace{-0.2cm}\int_0^{t^{\prime}} \hspace{-0.35cm} dt^{\prime \prime} R(t^{\prime},t^{\prime \prime})
\mathcal{V}^{\prime}(C(t,t^{\prime \prime}))+
 \label{NECttp}\\ 
&+\hspace{-0.15cm}\int_0^t \hspace{-0.25cm} dt^{\prime \prime} C(t^{\prime},t^{\prime \prime})
R(t,t^{\prime \prime})\mathcal{V}^{\prime \prime}(C(t,t^{\prime \prime}))  
+\beta\overline C(t^{\prime})\mathcal{V}^{\prime}(\overline C(t)) \nonumber \\
\partial_t R&(t,t^{\prime})+\mu(t)R(t,t^{\prime})=\hspace{-0.15cm} \int_{t^{\prime}}^t \hspace{-0.25cm} dt^{\prime \prime} R(t^{\prime \prime},t^{\prime})R(t,t^{\prime \prime})\mathcal{V}^{\prime \prime}(C(t,t^{\prime \prime})) 
\label{NERttp} \\
&\hspace{1cm}\overline{C}(t)=\beta \int_0^t \hspace{-0.25cm} dt^{\prime \prime} R(t,t^{\prime \prime}) \mathcal{V}^{\prime}(\overline C(t^{\prime \prime})) \ .
\label{NECbt}
\end{align}
\end{subequations}
Where $R(t,t^{\prime})=\frac{1}{N(1-c)}\sum_{i=cN}^{N}\left.\frac{\delta s_i(t)\rangle}{\delta h_i(t^{\prime})}\right |_{h_i=0}$
is the response function ($h_i(t^{\prime})$ is a magnetic field coupled to spin $i$) and the new correlation function $\overline C(t)$
is the correlation between the running  configuration at time $t$ and the pinned configuration $\overline{\mathcal{C}}$. 
By imposing the spherical constraint, one obtains the evolution equation for $\mu(t)$:\footnote{Note that 
the spherical parameter $\mu(t)$ is not related to the energy per spin in the same simple way found for the 
unconstrained p-spin spherical model.}
\begin{align}
\mu(t)&-T= \int_0^t \hspace{-0.25cm} dt^{\prime \prime} R(t,t^{\prime \prime}) 
\mathcal{V}^{\prime}(C(t,t^{\prime \prime}))+ \label{NEmut}\\
&+\int_0^t \hspace{-0.25cm} dt^{\prime \prime} C(t,t^{\prime \prime})R(t,t^{\prime \prime})
\mathcal{V}^{\prime \prime}(C(t,t^{\prime \prime}))+\beta\overline C(t)\mathcal{V}^{\prime}(\overline C(t)) \nonumber
\end{align}
The correlation function $\overline C(t)$ measures how far the system is from reaching equilibrium:
$\overline C(t)$ starts from zero at $t=0$ since an infinite temperature configuration is completely uncorrelated from one at equilibrium  
at temperature $T$. Afterwards, it increases and {\it if the system reaches equilibrium} then it becomes equal 
to the typical correlation between two equilibrium configurations, i.e. the self-overlap $q$ computed in the previous section, or equivalently 
the long $t-t'$ limit of $C(t,t')$, see top panel of Fig.3.  
An asymptotic limit of $\overline C(t)$ below $q$ instead indicates that the system is out of equilibrium even at long times. 
Actually, this corresponds to the so called
BIC test \cite{ANDREA}, which has been introduced in recent numerical studies on systems with pinned particles, where reaching equilibrium is a delicate issue. \\
Guided by the knowledge of the dynamical behavior at $c=0$, we expect a system quenched just slightly below the dynamical transition line to never equilibrate and to show the aging dynamics discovered in \cite{CK}. Indeed, our numerical solution\footnote{
We used the efficient numerical integration method proposed by Kim and Latz in~\cite{KL} and developed in~\cite{Kunietal}. The main idea of this method consists in using an adaptive integration step to better describe the short and long time shape of the correlation function.
} of~\eqref{NEeq} confirms this intuition (see Fig.\ref{NEdynAG}):
\begin{figure}
\hspace{-0.4cm}
\onefigure[width=6.1cm, angle=-90]{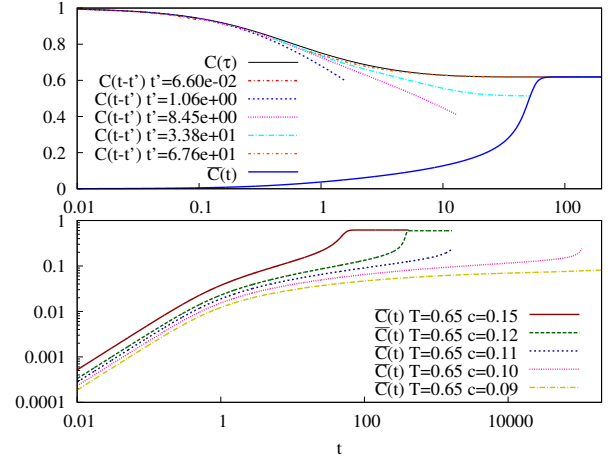}
\caption{Top panel: a successful BIC test showing that the asymptotic 
limit of the correlation function (as a function of $\tau=t-t'$) and $\overline C(t)$ have
the same large time value ($T$ and $c$ correspond to the point A3 of Fig.\ref{PhDi}). Bottom panel:
$\overline{C}(t)$ for several different concentrations of pinned spins
at $T=0.65$. The relaxation time rapidly grows approaching $c_{d_f}$
from the right.}
\label{NEdyn}
\end{figure}
$C(t,t^{\prime})$ strongly depends on the waiting time $t^{\prime}$ even for large time-scales,
but a plot of the same curves 
in terms of the relevant adimensional variable $\theta=t/t^{\prime}$ shows a good collapse.
At the same time, on time-scales of the order of one time translation invariance (TTI) and the fluctuation dissipation theorem (FDT) hold.
On this basis and following the analysis~\cite{CK} we 
simplify eqs.~\eqref{NEeq} by separately investigating 
the TTI regime for which $\tau=t-t^{\prime}$ such that $(t-t^{\prime})/t\rightarrow0$ and the aging regime for which  
$(t-t^{\prime})/t\sim O(1)$ and where we use $\theta=t^{\prime}/t$.
In the first regime the equation on $C(t,t^\prime)$ is related to the one of $R(t,t^{\prime})$ by FDT and reads:
\begin{align}
\partial_{\tau}C(\tau)+&\left(\beta\mathcal{V}^{\prime}(1)-\mu_{\infty}+\frac{1}{\beta}\right)(1-C(\tau))= \label{NECtau}\\
&-\frac{1}{\beta}C(\tau)-\beta\int_0^{\tau}du \partial_uC(u)\mathcal{V}^{\prime}(C(\tau-u)) \ .\nonumber
\end{align}
Here $\mu_{\infty}$ is the large time value of the spherical parameter in the non-equilibrium regime.
Taking the large $\tau$ limit of~\eqref{NECtau}, $\mu_{\infty}$ is found directly related to  
$C(\tau\rightarrow\infty)=r_1$:
\begin{equation}
\label{muinfr1}
\beta\mu_{\infty}=\frac{1}{1-r_1}+\beta^2\left[\mathcal{V}^{\prime}(1)-\mathcal{V}^{\prime}(r_1)\right] \ .
\end{equation}
To obtain the value of  $\mu_{\infty}$ 
one has to solve the aging regime since $\mu_{\infty}$ depends on it, see~\eqref{NEmut}.   
In order to do that, we first introduce the notation $C(t,t^{\prime})=r_1\mathcal{C}(\theta)$ and 
$R(t,t^{\prime})=\mathcal{R}(\theta)$. 
Since $\lim_{\tau\rightarrow\infty}C(\tau)=r_1$, we have $\mathcal{C}(1)=1$. The $t\gg t'$ limit of the correlation function
is denoted $r_0$, hence  $\mathcal{C}(0)=r_0/r_1$.
The two equations on $\mathcal{C}(\theta)$ and $\mathcal{R}(\theta)$, that we do not reproduce here, coincide under the hypothesis 
that $\mathcal{C}$ and $\mathcal{R}$ obey to a generalized FDT relation $\mathcal{R}(\theta)=x\beta r_1\mathcal{C}^{\prime}(\theta)$, where $x$ quantifies the FDT violation ($x=1$ means no violation).\\
 In summary, 
for a quench to a given point in the plane $c-T$ five parameters should be determined to characterize the aging behavior: $x, \mu_{\infty}, \overline C_{\infty}, r_1,$ and $r_0$. 
The five equations needed for the solution are \eqref{NECbt}, \eqref{NEmut}, \eqref{muinfr1}, 
and the two ones given by 
the equations on $\mathcal{C}(\theta)$ (or $\mathcal{R}$) evaluated in $\theta=1$ and $\theta=0$.
The final set of equations includes~\eqref{muinfr1} and
\begin{subequations}
\label{alleq}
\begin{align}
&\beta\mu_{\infty} =1+\label{muinfag}\\
&+\beta^2\left[\mathcal{V}^{\prime}(1)+\overline C_{\infty}\mathcal{V}^{\prime}(\overline C_{\infty})-
(1-x)r_1\mathcal{V}^{\prime}(r_1)-xr_0\mathcal{V}^{\prime}(r_0)\right] \nonumber\\
&\hspace{0cm}\overline C_{\infty} = \beta^2\mathcal{V}^{\prime}(\overline C_{\infty})(1-r_1(1-x)-xr_0)\label{Cbinfag}\\
&\hspace{0cm}1=\beta^2(1-r_1)^2\mathcal{V}^{\prime \prime}(r_1)
\label{marginalityeq}\\
&\beta\mu_{\infty}r_0=\beta^2\left[\mathcal{V}^{\prime}(r_0)+r_0\mathcal{V}^{\prime}(1)+\overline C_{\infty}\mathcal{V}^{\prime}(\overline C_{\infty})+\right.
\label{r0eq}\\
&\hspace{0.8cm}\left.-(1-x)r_1\mathcal{V}^{\prime}(r_0)-(1-x)r_0\mathcal{V}^{\prime}(r_1)-2xr_0\mathcal{V}^{\prime}(r_0)  \right] \ .\nonumber
\end{align}
\end{subequations}
We have studied the solution of these equations varying the values of $T$ and $c$. In order to avoid spurious results, 
we follow the solution starting from the dynamical transition line, where one can show that
$r_1=q_1$, $r_0=q_0=\overline C_{\infty}$, $x=1$, and 
$\mu_{\infty}=\mu_{\text{eq}}=1/\beta+\beta\mathcal{V}^{\prime}(1)$.
Starting from $c_d(T)$ and increasing $c$ or decreasing $T$, we find that $r_0$ and $\overline C_{\infty}$ rise gently, 
$x$ declines from one and $r_1$ remains similar to $q_1$ (in general $r_0<\overline C_{\infty}\ll r_1$).
For a large enough value of $c$ (or a small enough value of $T$) the solution ceases to exist. This leads to 
a second dynamical transition line $T_{d_f}(c)$, or equivalently $c_{d_f}(T)$, that meets 
the MCT dynamical transition line at $(c_h,T_h)$, see Fig.2. 
\begin{figure}
\onefigure[width=5.5cm, angle=-90]{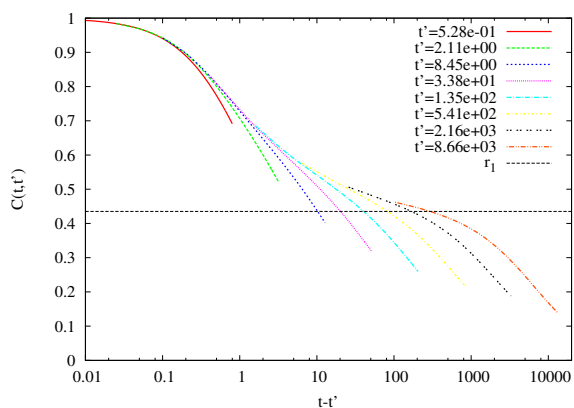}
\caption{Aging behavior of $C(t,t^{\prime})$ for a quench at the point A2 of Fig.\ref{PhDi}. 
The dotted line denotes the plateau value $r_1$.}
\label{NEdynAG}
\end{figure} 
Within the region delimited by the two dynamical transition lines the system never equilibrates and ages forever. 
In this regime the physical behavior is the one already understood for the unconstrained p-spin model. Actually, the free
energy landscape of the randomly pinned p-spin spherical model is very similar to the one of the unconstrained 
system, except that there is one special state that is singled out and favored by the random pinning field \cite{BC}. 
After a quench between $c_d(T)$ and $c_{d_f}(T)$ the system ages because in its descent 
in the free-energy landscape, it is never able to drift lower than the
so called marginal states. These "trap" the system: the closer it approaches them, the fewer are the 
unstable modes allowing it to escape. The existence of a preferred state correlated with the random pinning field 
is irrelevant since the system is not able to visit it in this regime.
In order to be sure that this interpretation is correct we 
also independently obtained~\eqref{muinfr1}, \eqref{muinfag}, \eqref{Cbinfag}, and \eqref{r0eq}
using the replica formalism. In this context $C_{\infty}$ is the overlap between marginal states and equilibrium states,
$r_1$ and $r_0$ represents the self and the mutual overlap of marginal metastable states and 
$x$ is the block parameter of the one step replica symmetry breaking matrix. \\
Let us now discuss the nature of the transition at $c_{d_f}(T)$, which is different from the usual MCT one.
In this case the existence of a preferred state makes a difference. 
In fact, beyond $c_{d_f}(T)$ the pinning field is so strong that the marginal states disappear and the system 
ends up in the equilibrium state favored by the field. However, contrary to what happens at $c_d(T)$ 
the marginal states are not typical equilibrium states, actually they are quite different from those. Thus, 
approaching $c_{d_f}(T)$ from the right the system ages for a certain time, since it is lost around what remains 
of the marginal states, and then at certain point it finds some escaping direction and
relaxes to equilibrium. Thus, as it happens for quenches close to the MCT transition line, aging eventually is interrupted
but in a way which is very different.  
This is clearly shown in the lower panel of Fig.\ref{NEdyn}: 
$\overline C(t)$ first rises toward an intermediate value, then it suddenly jumps
to the equilibrium self-overlap. Concomitantly, $C(t,t^{\prime})$ stops to age and TTI and FDT start to hold. 
Approaching $c_{d_f}(T)$ the time to reach equilibration becomes
longer and longer, possibly diverging as a power law in $c-c_{d_f}(T)$. 
In order to understand better the nature of this transition it is useful to introduce two different
timescales: the equilibrium decorrelation time, $\tau_{eq}$, which is the time it takes to the equilibrium correlation 
function $C(\tau)$ to decrease to its asymptotic limit and the relaxation time, $\tau_{rel}$, which is the time it 
takes to $\overline C(t)$
to reach the self-overlap equilibrium value. Approaching $c_d(T)$ from the left, $\tau_{eq}$ and $\tau_{rel}$
both diverge. Instead, approaching $c_{d_f}(T)$ from the right only $\tau_{rel}$ diverges: the system takes a long time 
to find the way to escape from (almost) marginal states but then, when this is found, the system relaxes to the equilibrium state which is very different
from the marginal states and within which relaxation is fast. 
This kind of transition has been also found in some quantum spin disordered models \cite{QuantumLeticia} and dubbed 
first order dynamical phase transition. We think that physically it resembles more to a dynamic spinodal because it is related to 
the sudden disappearance of the marginal states. 

\section{Discussion and Conclusion}
The two main results of the previous sections, based on mean field theory, concern the 
MCT phase diagram for randomly pinned glassy systems and the aging behavior for quenches in the $c-T$ plane. 
Let us now discuss them, especially their extension to finite dimensions and realistic systems. 
First, we would like to stress that although our results are obtained for an admittedly very abstract model, they are direct consequences of the evolution of the free-energy landscape in the $c-T$ plane. Thus, they are expected to be valid  
for all systems sharing a similar landscape and in particular, within RFOT theory, 
also for particle models in finite dimensions.
As anticipated in the introduction our MCT phase diagram differs from the one
obtained for hard spheres by Krakoviack \cite{KRAKO} using a generalization of the MCT projection operators method.
He finds that the transition line does not end in a critical point. Instead, it goes on but the transition 
becomes continuous. 
On the contrary, after the end-point, we just find a Widom cross-over line where the decorrelation time is maximum but not infinite\footnote{
Single particle Lorentz gas physics and percolation phenomena may play a role at high temperature, where higher concentrations of pinned particles are required to reach the interesting region of the phase diagram. 
These could make the observation of collective dynamical slowing down more difficult.}.
At this stage it is not clear what is the origin of this difference, besides the fact that our approach only focuses on collective
dynamical slowing down whereas Karkoviack's one somehow put together single-particle and many particles phenomena. 
However, this disagreement between the two approaches 
partially fades out when the results are transposed in finite dimensions since the MCT transition line becomes a cross-over anyway. It 
would be interesting to perform either the replicated HNC computation of \cite{FCP} or the cage expansion of \cite{MP,PZ} for randomly pinned systems to better clarify this issue and obtain quantitative results complementary    
to the ones in \cite{KRAKO}.  \\
Let us now focus on the dynamical behavior below $T_d(c)$. 
Since the MCT transition is avoided in finite dimensions, $\tau_{rel}$ only diverges at $T_K(c)$ and in a
a super-Arrhenius way. Increasing $c$ the divergence of the relaxation time becomes milder. Indeed at $(c_h,T_h)$ the exponent of the generalized Vogel-Fulcher-Tamman law becomes equal to $1/6$ within mean-field theory (and 
$1$ within the Migdal-Kadanoff RG procedure of \cite{BC}) while it is equal to $2$ at $c=0$ \cite{BC}. Thus, the system becomes stronger and, as discussed previously, less dynamically heterogeneous.      
This is in agreement with the results of numerical simulations \cite{KMS}. We now consider the 
aging dynamics. 
For an unconstrained system, quenches above $T_K(0)$ 
lead to interrupted aging whereas quenches below it lead to
aging that goes on forever. Instead, for constrained ones, we expect interrupted aging also below $T_K(c)$ 
(in particular above $T_{d_f}(c)$). The reason relies in the form of the free energy landscape: after the 
quench the system is trapped and ages on high free-energy states in way very similar to what happens 
above $T_K(c)$. However, below $T_K(c)$, the state 
correlated with the pinning field is thermodynamically favored with respect to all the others even taking into account their multiplicity.
Thus, when the waiting time becomes of the order of the nucleation time of the favored state the system stops aging and 
suddenly\footnote{Actually the nucleation process can be slow because it takes place in a random environment.} reaches equilibrium\footnote{Remarkably, a similar phenomenon plays an important role in optimization problems such as  compressed sensing\cite{CS}.}, accordingly the decorrelation time within the equilibrium state is very different from the relaxation time,  the former is expected to remain small whereas the latter diverges approaching  
$T_K(c)$ from below. This is relevant for the so called BIC test: it means that 
in this regime starting from the pinned configurations the system is automatically at equilibrium and decorrelates rapidly
\footnote{However, microscopic times
could be large in presence of pinned particles.}. Thus 
one can easily obtain equilibrated averages even though the time it takes to get a successful BIC test could be very large\footnote{The problem in a simulation is that one cannot be sure to be below $T_K(c)$.}. \\
Numerical simulations and even experiments in colloids using optical trapping techniques can be 
performed to test our predictions. In particular, it would be very interesting to observe the nucleation-interrupted aging
below $T_K(c)$ and the characteristic dynamical features taking place near the spinodal point $(c_h,T_h)$. 
\acknowledgments
We thank L.F. Cugliandolo and K. Miyazaki for useful remarks, A. Lefevre and K. Miyazaki for help with the 
numerical code and V. Krakoviack for sharing with us unpublished results on randomly pinned hard spheres. 
We acknowledge support from the ERC grant NPRGGLASS.

\end{document}